\RequirePackage{fix-cm}
\documentclass[smallextended]{svjour3}       
\smartqed  
\usepackage{graphicx}
\newcommand{\ket}[1]{\left| #1 \right\rangle}
\newcommand{\bra}[1]{\left\langle #1 \right|}
\newcommand{\Tr}{\text{Tr}}
\usepackage[breaklinks=true,colorlinks=true,linkcolor=blue,urlcolor=blue,citecolor=blue]{hyperref}
\usepackage{enumerate}
\usepackage{mathtools}
\usepackage{amsfonts}
\usepackage{amssymb}
\usepackage{color}
\usepackage{cite}
\usepackage{float}

\begin{document}

\title{Entanglement fidelity and measure of entanglement 
}


\author{Vahid Azimi Mousolou        
}


\institute{V. Azimi Mousolou \at
              Department of Applied Mathematics and Computer Science, Faculty of Mathematics and Statistics, University of Isfahan, Box 81745-163 Isfahan, Iran. \\
              School of Mathematics, Institute for Research in Fundamental Sciences (IPM), P. O. Box 19395-5746, Tehran, Iran.\\
              Department of Physics and Astronomy, Uppsala University, Box 516, Se-751 20 Uppsala, Sweden.\\
              Tel.: +9837934641\\
              \email{v.azimi@sci.ui.ac.ir}           
}

\date{Received: date / Accepted: date}

\maketitle

\begin{abstract}
The notion of entanglement fidelity is to measure entanglement preservation through quantum channels. Nevertheless, the amount of entanglement present in a state of a quantum system at any time is measured by quantities known as measures of entanglement. Since there are different types of measures of entanglement, one may expect an entanglement fidelity to associate with its own measure of entanglement counterpart. Here, we aim to investigate association between the so called entanglement fidelity and some measures of entanglement, namely, entanglement of formation, concurrence and negativity. New entanglement fidelities based upon these measures of entanglement are introduced and statistically compared with the so called previously introduced entanglement fidelity. It is shown that the entangling aspect of the so called entanglement fidelity is neither of type entanglement of formation and concurrence nor of type negativity. The results, in addition, expose inability of the so called entanglement fidelity for detecting, in a broad sense, entanglement preservation through quantum channels. Our analyses opens up a new venue in the study of entanglement fidelity and measure of entanglement by demonstrating that each measure of entanglement solely defines its own entanglement fidelity.
\end{abstract}

\section{Introduction}
Entanglement as one of the main notion of quantum source of information has been always at the centre of attention in quantum sciences and technologies. 
The fundamental roles of quantum entanglement in quantum cryptography, superdense coding, quantum teleportation,
quantum error correction, efficient quantum computation and many other applied and basic quantum sciences \cite{steane1998, bennet2000, horodecki2009, niellsen2010}, have turned the study of entanglement into a major area of research.  

Concerning the concept of quantum entanglement, the two relevant questions that naturally arise are: how to quantify and compare entanglement in quantum states? and how well entanglement of a quantum state is maintained and preserved through quantum channels during a quantum information processing? Although these questions have been addressed extensively in many research works, there are still much that remain to be explored. For the first question we encounter a concept known as measure of entanglement and the latter question leads us to a concept known as entanglement fidelity. So far different classes of measures of entanglement, such as entanglement of formation, concurrence, entanglement of distillation, relative entropy of entanglement, negativity, Bures metric, geometric measure of entanglement, etc., have been introduced. For a review about measures of entanglement one may see Re. \cite{horodecki2009}.  The quantity of entanglement fidelity, introduced and discussed in pioneering works \cite{schumacher1996,nielsen1996,barnum1998}, is believed to provide a measure of how successfully the entanglement between a pair of  quantum subsystems would be preserved through a quantum process.

Despite that all measures of entanglement follow the same criteria \cite{vedral1997, vedral1998, vidal2000}, they do not all impose the same ordering in a set of states \cite{virmani2000, miranowicz2004}. In a word, not all the measures of entanglement behave mathematically in a same manner and this may imply that there are different types of entanglements present in a quantum system. From this point of view, as the entanglement fidelity measures entanglement preservation in a quantum state going through a quantum process, it is natural to ask what type of entanglement is of concern in the entanglement fidelity. To study correlations between the entanglement fidelity and measures of entanglement, here we consider some measures of entanglement, namely, the entanglement of formation, concurrence \cite{hill1997, wootters1998} and the negativity \cite{vidal2002}, which have been shown to quantify different aspects of entanglement \cite{miranowicz2004}. In fact, we investigate ordinal correlations between the so called entanglement fidelity discussed in Refs. \cite{schumacher1996, nielsen1996, barnum1998} and these measures of entanglement. To this end, associated with entanglement of formation, concurrence and negativity we introduce fidelity type quantities and statistically compare them with the well known entanglement fidelity using Kendall rank correlation coefficient \cite{kendall1990}. 
Our analyses demonstrate that each measure of entanglement specifies its own entanglement fidelity and the entanglement fidelity in Refs. \cite{schumacher1996, nielsen1996, barnum1998} is not related to any of the measures of entanglement, which the present work concerns. Moreover, we show that in some cases, the entanglement fidelity in Refs. \cite{schumacher1996, nielsen1996, barnum1998} is not able to detect entanglement preservation through quantum channels.
We notice that the entangling aspect of the entanglement fidelity has been questioned before from different perspective and approach \cite{xiang2007}. 

The paper is organized as it follows: In sec. \ref{Entanglement Fidelity}, the entanglement fidelity based on Refs. \cite{schumacher1996, nielsen1996, barnum1998} is briefly reviewed. We discuss entanglement of formation and concurrence in sec. \ref{Entanglement of Formation and Concurrence} and define associated fidelity type quantities. We recall the negativity and introduce an associated fidelity type quantity in sec. \ref{Negativity}. In sec. \ref{Data Analysis and Discussion}, 
we examine entangling aspect of the entanglement fidelity in Refs. \cite{schumacher1996, nielsen1996, barnum1998} 
by performing data analyses and discussing ordinal correlations between the entanglement fidelity and the fidelities associated with entanglement of formation, concurrence and negativity. The paper ends with a summary in sec. \ref{Summary}.

\section{Entanglement Fidelity}
\label{Entanglement Fidelity}
 In this section we briefly recall the quantity of entanglement fidelity based on Refs. \cite{schumacher1996, nielsen1996, barnum1998}.

Consider a quantum system of combined two quantum subsystems labeled as $R$ and $Q$. Suppose the joint system $RQ$ initially is prepared in a general pure state $\rho_{\text{i}}^{RQ}=\ket{RQ}\bra{RQ}$. Further assume that the subsystem $Q$ undergoes some evolutions described by a quantum operation $\mathcal{E}$ while the subsystem $R$ is dynamically isolated. In this case, the overall dynamics of the joint system $RQ$ is described by the quantum operation $I\otimes\mathcal{E}$, where $I $ here is the identity operator acting on the subsystem $R$. Thus the final state of the joint system is given by the density operator $\rho_{\text{f}}^{RQ}=I\otimes\mathcal{E}(\rho_{\text{i}}^{RQ})$. For such a process, the quantity 
\begin{eqnarray}
F_{\text{e}}=<\rho_{\text{i}}^{RQ},\rho_{\text{f}}^{RQ}>_{\text{HS}}=\Tr(\rho_{\text{i}}^{RQ}\rho_{\text{f}}^{RQ})=\bra{RQ}\rho_{\text{f}}^{RQ}\ket{RQ}\nonumber\\
\label{EF}
\end{eqnarray}
defined as the Hilbert-Schmidt inner product between two states $\rho_{\text{i}}^{RQ}$ and $\rho_{\text{f}}^{RQ}$, is believed to quantify the entanglement fidelity indicating the variation of the entanglement in the quantum process \cite{schumacher1996, nielsen1996, barnum1998}. The $F_{\text{e}}$, in fact, takes its value in the interval $[0,1]$, where values close to $1$ are supposed to imply that the entanglement is well preserved and values close to $0$ indicate that the entanglement is mostly destroyed. 

Although the quantity $F_{\text{e}}$ given in Eq. (\ref{EF}) is the state fidelity (squared) between joint initial state $\rho_{\text{i}}^{RQ}$ and final state  $\rho_{\text{f}}^{RQ}$, it has been shown in Ref. \cite{schumacher1996} that $F_{\text{e}}$ actually is an intrinsic property of the subsystem $Q$ itself and depends solely on the initial state of the subsystem $Q$ given by the reduced density operator 
\begin{eqnarray}
\rho_{\text{i}}^{Q}=\Tr_{R}\rho_{\text{i}}^{RQ},
\end{eqnarray}
where $\Tr_{R}$ indicates partial trace over the subsystem $R$, and the quantum channel $\mathcal{E}$ to which the subsystem $Q$ is subjected. It has been further shown in 
Ref. \cite{schumacher1996} that the $F_{\text{e}}$ does not in general agree with the state fidelity (squared) between initial and final states of the subsystem $Q$, i.e. $F(\rho_{\text{i}}^{Q}, \rho_{\text{f}}^{Q})=\Tr(\rho_{\text{i}}^{Q}\rho_{\text{f}}^{Q})$, where $\rho_{\text{\text{f}}}^{Q}=\Tr_{R}\rho_{\text{f}}^{RQ}$. In fact, the following general relation holds 
\begin{eqnarray}
F_{\text{e}}\equiv F_{\text{e}}(\rho_{\text{i}}^{Q}, \mathcal{E} )\leqslant F(\rho_{\text{i}}^{Q}, \rho_{\text{f}}^{Q}).
\end{eqnarray}

An interesting question, which may arise here, is if $F_{\text{e}}$ and $F$  are both kind of state fidelities depending only on the initial state and the quantum channel, why $F_{\text{e}}$ and $F$ do not in general agree? Is it simply because of the mathematical fact that the two operators $\Tr_{R}$ and $\mathcal{E}$ do not in general commute, i.e., $\Tr_{R}(I\otimes \mathcal{E}(\rho_{\text{i}}^{RQ}))\ne \mathcal{E}(\Tr_{R}(\rho_{\text{i}}^{RQ}))$, or $F_{\text{e}}$ and $F$ are in principle related to different quantum concepts? It is argued in Refs. \cite{schumacher1996, nielsen1996, barnum1998} that $F_{\text{e}}$ is related to how well the quantum entanglement between two subsystems $R$ and $Q$ present in the state $\rho_{\text{i}}^{RQ}$ is preserved through the quantum process $\mathcal{E}$ while the $F$ is a useful distance measure of how far apart the two state $\rho_{\text{i}}^{Q}$ and $\rho_{\text{f}}^{Q}$ are.

However, in some cases we notice that the $F_{\text{e}}$ lacks detecting true entanglement preservation or variation in a quantum process. For instance, an initial two-qubit Bell state can be mapped to a different Bell state with a purely local Pauli operator acting only on one of the qubit subsystems. Since both initial and final Bell states are maximally entangled, this process corresponds to a well entanglement preserving quantum process, but from the orthogonality of two different Bell states we have $F_{\text{e}}=0$, which indicate complete loss of entanglement. This contradicts what we stated above about the quantum nature of $F_{\text{e}}$. Nevertheless, one may still argue that since there are different measures of entanglement or in some sense different types or different aspects of entanglement, the quantity  $F_{\text{e}}$ must be investigated in a broader scale to see what type of entanglement the $F_{\text{e}}$ concerns if there is any. Below we examine correlations between the fidelity $F_{\text{e}}$ and two types of entanglement given by concurrence, a measure of the entanglement of formation, and negativity, a measure of the entanglement cost.

\section{Entanglement of Formation, Concurrence and associated Fidelities}
\label{Entanglement of Formation and Concurrence}
One of the fundamental measure of entanglement, which is in some sense defined based on the amount of resources needed to form a given entangled state, is known as entanglement of formation \cite{bennett1996}. An explicit mathematical formulation of the entanglement of formation for a pair of qubits has been established in Refs. \cite{hill1997, wootters1998}. This explicit formula is given in terms of a quantity called concurrence \cite{hill1997, wootters1998}, which on its own introduces a measure of entanglement as well. These measures are defined as follows.

For a given mixed  state density operator $\rho^{RQ}$ of a pair of quantum subsystems $R$ and $Q$, the entanglement of formation is defined as \cite{bennett1996}
\begin{eqnarray}
E(\rho^{RQ})=\min\sum_{k}p_{k}E(\ket{\psi_{k}}),
\label{EofF}
\end{eqnarray}
where $E(\ket{\psi_{k}})$ is the pure state entanglement given by the von Neumann's entropy of either of the two subsystems  $R$ and $Q$, i.e.,
\begin{eqnarray}
E(\ket{\psi_{k}})=-\Tr(\rho_{k}^{R}\log_{2}\rho_{k}^{R})=-\Tr(\rho_{k}^{Q}\log_{2}\rho_{k}^{Q}),
\end{eqnarray}
for reduced density operators 
$\rho_{k}^{R}=\Tr_{Q}(\ket{\psi_{k}}\bra{\psi_{k}})$ and $\rho_{k}^{Q}=\Tr_{R}(\ket{\psi_{k}}\bra{\psi_{k}})$. The minimum in Eq. (\ref{EofF}) is taken over all possible pure-state decompositions of 
\begin{eqnarray}
\rho^{RQ}=\sum_{k}p_{k}\ket{\psi_{k}}\bra{\psi_{k}}.
\end{eqnarray}

In two qubits case, the entanglement of formation can be explicitly expressed as a computable function of  the state density operator $\rho^{RQ}$ \cite{hill1997, wootters1998}. This computable mathematical description make use of the spin flip transformation, which for general two-qubit mixed state $\rho^{RQ}$ reads 
\begin{eqnarray}
\tilde{\rho}^{RQ}=\sigma_{y}\otimes\sigma_{y}\bar{\rho}^{RQ}\sigma_{y}\otimes\sigma_{y}.
\end{eqnarray}
Here $\bar{\rho}^{RQ}$ is the complex conjugate of $\rho^{RQ}$ taken in the standard two-qubit computational basis $\{\ket{00}, \ket{01}, \ket{10}, \ket{11}\}$, and 
\begin{eqnarray}
\sigma_{y}=
\left(
\begin{array}{cc}
 0 &  -i   \\
  i & 0      
\end{array}
\right)
\label{pauli2}
\end{eqnarray}
is the second component of Pauli matrices in the single-qubit computation basis $\{\ket{0}, \ket{1}\}$. Considering $\lambda_{1}, . . . ,\lambda_{4}$ to be the eigenvalues of the hermitian matrix $\sqrt{\sqrt{\rho^{RQ}}\tilde{\rho}^{RQ}\sqrt{\rho^{RQ}}}$ in decreasing order, the entanglement of formation of the two-qubit mixed state $\rho^{RQ}$ can be written as \cite{wootters1998}
\begin{eqnarray}
E(\rho^{RQ})=-\xi\log_{2}\xi-(1-\xi)\log_{2}(1-\xi),
\label{EoF}
\end{eqnarray}
where $\xi=\frac{1+\sqrt{1-[C(\rho^{RQ})]^{2}}}{2}$ for the concurrence $C(\rho^{RQ})$ defined as 
\begin{eqnarray}
C(\rho^{RQ})=\max\{0, \lambda_{1}-\lambda_{2}-\lambda_{3}-\lambda_{4}\}.
\end{eqnarray}
As mentioned above the concurrence $C(\rho^{RQ})$ by itself is also identified as a measure of entanglement \cite{wootters1998}. 

Associated with the above two measures of entanglement, entanglement of formation and concurrence, we may consider the following fidelity type quantities
\begin{eqnarray}
F_{\text{ef}}=1-|E(\rho^{RQ})-E(I\otimes\mathcal{E}(\rho^{RQ}))|\nonumber\\
F_{\text{c}}=1-|C(\rho^{RQ})-C(I\otimes\mathcal{E}(\rho^{RQ}))|.
\label{FEFC}
\end{eqnarray}
Similar to the entanglement fidelity $F_{\text{e}}$, these quantities also have their values in the interval $[0, 1]$ and allow us to evaluate how the amount of entanglement quantified by entanglement of formation or concurrence are preserved, when the system is subjected to the quantum channel $I\otimes\mathcal{E}$. The values of $F_{\text{ef}}$ and $F_{\text{c}}$ close to $1$ indicate that the entanglement of formation and concurrence are well preserved, and the values close to $0$ imply that the entanglement of formation and concurrence are mainly lost. In fact, $F_{\text{ef}}$ and $F_{\text{c}}$ can be, respectively, regarded as fidelity of entanglement of formation and fidelity of concurrence in a quantum process, where the subsystem $Q$ undergoes some evolutions described by a quantum operation $\mathcal{E}$ while the subsystem $R$ is dynamically isolated. 

\section{Negativity and associated Fidelity}
\label{Negativity}
Another measure of entanglement that we consider here is known as negativity. Negativity, which in a sense measures the entanglement cost of a quantum state \cite{vidal2002}, can be regarded as a quantitative version of the Peres-Horodecki criterion \cite{peres1996, horodecki1996}. Although, in the case of two-qubit pure states, negativity coincide with concurrence, the two measures of entanglement behave very differently in general \cite{miranowicz2004} and are believed to reflect different types of entanglement present in a quantum system. The negativity for a general bipartite mixed state $\rho^{RQ}$ reads \cite{vidal2002, miranowicz2004}
\begin{eqnarray}
N(\rho^{RQ})=\|[\rho^{RQ}]^{T_{R}}\|_{1}-1,
\label{Ne}
\end{eqnarray}
where $T_{R}$ denotes the partial transpose with respect to subsystem $R$ and thus 
\begin{eqnarray}
\bra{i_{R}j_{Q}}[\rho^{RQ}]^{T_{R}}\ket{k_{R}l_{Q}}=\bra{k_{R}j_{Q}}\rho^{RQ}\ket{i_{R}l_{Q}}
\end{eqnarray}
for a given orthonormal product basis $\ket{i_{R}j_{Q}}=\ket{i_{R}}\otimes\ket{j_{Q}}\in\mathcal{H}_{R}\otimes\mathcal{H}_{Q}$. For any operator $A$, $\|A\|_{1}=\Tr\sqrt{A^{\dagger}A}$ is the trace norm, which is equal to the sum of the absolute values of the eigenvalues of $A$ in the case of hermitian operator $A$. Since $\Tr[[\rho^{RQ}]^{T_{R}}]=1$, the negativity in Eq. (\ref{Ne}) is actually twice the sum of the absolute values of the negative eigenvalues of $[\rho^{RQ}]^{T_{R}}$ \footnote{Note that we here use the negativity defined in Ref. \cite{miranowicz2004}, which is twice the negativity introduced in Ref. \cite{vidal2002}.}.

Similar to the previous section, we may consider the quantity, 
\begin{eqnarray}
F_{\text{n}}=1-|N(\rho^{RQ})-N(I\otimes\mathcal{E}(\rho^{RQ}))|,
\label{FN}
\end{eqnarray}
in order to evaluate the negativity type of entanglement fidelity or in short the fidelity of negativity. Indeed, the $F_{\text{n}}$ provides a measure of how well the negativity between subsystems $R$ and $Q$ is preserved by the quantum process $\mathcal{E}$. Here also we have $F_{\text{n}}\in[0, 1]$, where the values close to $1$ or $0$, respectively, indicate that the negativity is mainly preserved or lost. 

It is  shown in Ref. \cite{miranowicz2004} that concurrence and negativity do not impose the same ordering in a set of states, which may in a sense imply these two measures of entanglement refer independently to different types of entanglement. Therefore $F_{\text{c}}$ and $F_{\text{n}}$ are independent quantities and quantify different aspects of entanglement. 

\section{Mutual Correlations}
\label{Data Analysis and Discussion}
Having introduced the fidelities $F_{\text{e}}$, $F_{\text{ef}}$, $F_{\text{c}}$, and $F_{\text{n}}$, in previous sections, here we examine correlations between $F_{e}$ and the other fidelities to see if the entangling aspect of $F_{\text{e}}$ is the entanglement of formation and concurrence type or the negativity type. 

Consider a normalized general two-qubit pure state $\rho^{RQ}_{\text{i}}=\ket{RQ}\bra{RQ}$, where  
\begin{eqnarray}
\ket{RQ}=\alpha\ket{00}+\beta\ket{01}+\gamma\ket{10}+\delta\ket{11},\ \ \ \ \ \  \alpha, \beta, \gamma, \delta\in\mathbb{C}.
\label{samplestate}
\end{eqnarray}
Assume the initial state $\rho^{RQ}_{\text{i}}$ going through the local bit-phase flip channel given by the quantum operation $I\otimes \sigma_{y}$, where $\sigma_{y}$ is the Pauli operator in Eq. (\ref{pauli2}).
As $\sigma_{y}$ is only affecting the second qubit, the final state reads $\rho^{RQ}_{\text{f}}=\ket{\widetilde{RQ}}\bra{\widetilde{RQ}}$ with
\begin{eqnarray}
\ket{\widetilde{RQ}}=-\beta\ket{00}+\alpha\ket{01}-\delta\ket{10}+\gamma\ket{11}.
\end{eqnarray} 
Since the final vector state $\ket{\widetilde{RQ}}$ is orthogonal to the initial vector state $\ket{RQ}$, the $F_{e}$ vanishes, indicating that the entanglement is totally destroyed through this quantum channel. However,  
concurrence and negativity for initial and final states \footnote{For two-qubit pure states, concurrence and negativity are the same \cite{miranowicz2004}.}
 \begin{eqnarray}
C(\rho^{RQ}_{\text{i}})=N(\rho^{RQ}_{\text{i}})=2|\alpha\delta-\beta\gamma|=N(\rho^{RQ}_{\text{f}})=C(\rho^{RQ}_{\text{f}}),
\end{eqnarray} 
result $F_{\text{ef}}=F_{\text{c}}=F_{\text{n}}=1$, confirming perfect entanglement preservation through the quantum channel. Therefore, this example not only demonstrates the failure of the quantity $F_{e}$ in detecting entanglement 
preservation through the quantum process but also provides an evidence for $F_{e}$ being independent of entanglement of formation, concurrence and negativity. To clarify these independencies, below we further explore ordinal correlations between fidelities.

Note that from Eq. (\ref{EoF}), we have the entanglement of formation as an increasing function of concurrence. 
This implies that any ordinal correlation between $F_{\text{e}}$ and $F_{\text{c}}$ will hold true between $F_{\text{e}}$ and $F_{\text{ef}}$ as well and vice versa. 
Therefore in the following we only focus on the ordinal correlation
among the three quantities $F_{\text{e}}$, $F_{\text{c}}$ and $F_{\text{n}}$.

We employ the statistical tool known as Kendall rank correlation coefficient or in short Kendall's tau coefficient, which measures ordinal associations between two observed quantities \cite{kendall1990}. We evaluate the fidelities $F_{\text{e}}$, $F_{\text{c}}$, and $F_{\text{n}}$ for a specific family of two-qubit states subjected to some quantum noise channels. We then compare the collected data sets related to these fidelities pairwise for ordinal correlations via Kendall's tau coefficient.

\subsection{Kendall's tau coefficient }
For a set of paired observations $\{(x_{i}, y_{i})\}_{i=1}^{n}$ of two real-valued quantities $X$ and $Y$, the Kendall's tau coefficient is defined as 
\begin{eqnarray}
\tau=\frac{2}{n(n-1)}\sum_{i<j}\text{sgn}(x_{i}-x_{j})\text{sgn}(y_{i}-y_{j}),
\label{tau}
\end{eqnarray}
where sgn denotes the sign function.  Note that the coefficient is in the range $-1\le\tau\le 1$, and has the following properties
\begin{itemize}
\item If $\tau=1$, the two quantities $X$ and $Y$ perfectly follow the same ordering, i.e., 
\begin{eqnarray}
x_{i}\geqslant x_{j}\Longleftrightarrow y_{i}\geqslant y_{j},
\end{eqnarray}
 and thus there exist a direct correlation between the two quantities .
\item If $\tau=-1$ the two quantities $X$ and $Y$ perfectly follow the opposite ordering, i.e., 
\begin{eqnarray}
x_{i}\geqslant x_{j}\Longleftrightarrow y_{i}\leqslant y_{j},
\end{eqnarray}
 and thus there exist a reverse correlation between the two quantities.
\item The two quantities $X$ and $Y$ are independent if the Kendall's tau coefficient is approximately zero.
\item If $|\tau|\ne1$, the two quantities $X$ and $Y$ do not in principle  follow certain correlated ordinal patterns. Therefore, in this case, $X$ and $Y$ cannot both refer to some physical observations with the same physical properties and nature at each instance.    
\end{itemize}

\subsection{Data analysis and discussion}
We focus on two-qubit model systems prepared initially in general two-qubit pure state $\rho^{RQ}_{\text{i}}=\ket{RQ}\bra{RQ}$ given by Eq. (\ref{samplestate}).
We assume the qubit $Q$ is subjected to some quantum noise channels. The channels that we have in mind are ''amplitude damping'', 'bit flip''
and ''phase flip'' channels, which are described by the following operation elements 
\begin{itemize}
\item 
Amplitude damping:
\begin{eqnarray}
\left(
\begin{array}{cc}
 1 &  0   \\
  0 & \sqrt{1-p}      
\end{array}
\right),\ \ \ \ \ 
\left(
\begin{array}{cc}
 0 &  \sqrt{p}   \\
 0 & 0     
\end{array}
\right),
\label{Amplitude damping}
\end{eqnarray}
\item Bit flip: 
\begin{eqnarray}
\sqrt{1-p}\left(
\begin{array}{cc}
 1 &  0   \\
  0 & 1      
\end{array}
\right),\ \ \ \ \ 
\sqrt{p}\left(
\begin{array}{cc}
 0 &  1   \\
 1 & 0      
\end{array}
\right)
\label{Bit flip}
\end{eqnarray}
\item 
Phase flip:
\begin{eqnarray}
\sqrt{1-p}\left(
\begin{array}{cc}
 1 &  0   \\
  0 & 1      
\end{array}
\right),\ \ \ \ \ 
\sqrt{p}\left(
\begin{array}{cc}
 1 &  0   \\
 0 & -1     
\end{array}
\right),
\label{Phase flip}
\end{eqnarray}
\end{itemize}
where $p\in[0,1]$.

To pursue data collection and analysis, a randomly selected initial joint two-qubit state of type Eq. (\ref{samplestate}) is sent through one of the quantum channels introduced above for $M=2\times 10^{2}$ different values of $p$ uniformly distributed in $[0, 1]$. We emphasis that in each channel only the qubit $Q$ is affected by quantum noise operations and the qubit $R$ is left alone. For each value of $p$ the fidelities $F_{\text{e}}$, $F_{\text{c}}$, and $F_{\text{n}}$ are evaluated to produce corresponding three M-data sets. Then the M-data sets are mutually compared by Kendall's tau coefficient to detect ordinal correlations among the fidelities $F_{\text{e}}$, $F_{\text{c}}$, and $F_{\text{n}}$. This procedure is repeated for $\tilde{M}=5\times 10^{3}$ numbers of normally distributed random initial states of type Eq. (\ref{samplestate}). The results are as follow.

\begin{figure}[h]
\begin{center}
\includegraphics[width=80mm]{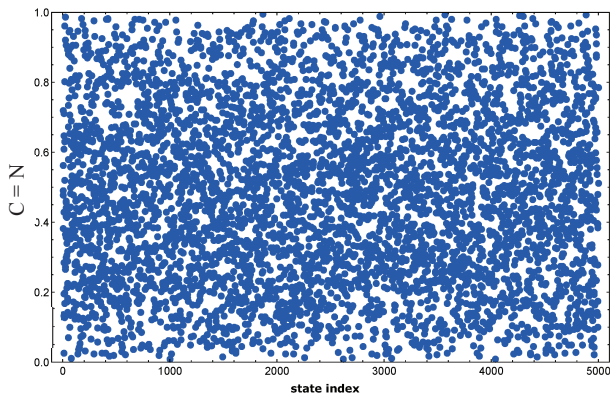}
\end{center}
\caption{(Color online). The concurrence and negativity distributions of the randomly selected sample of $\tilde{M}=5\times 10^{3}$ numbers of initial two-qubit pure states. For a two-qubit pure state, concurrence and negativity are the same.}
\label{fig:icnd}
\end{figure}

We use the same sample of initial states throughout the analysis. The concurrence and negativity distributions of the randomly selected sample of initial states are shown in Fig. \ref{fig:icnd}. As the initial states are two-qubit pure states, concurrence and negativity distributions are the same.

\begin{figure}[h]
\begin{center}
\includegraphics[width=85mm]{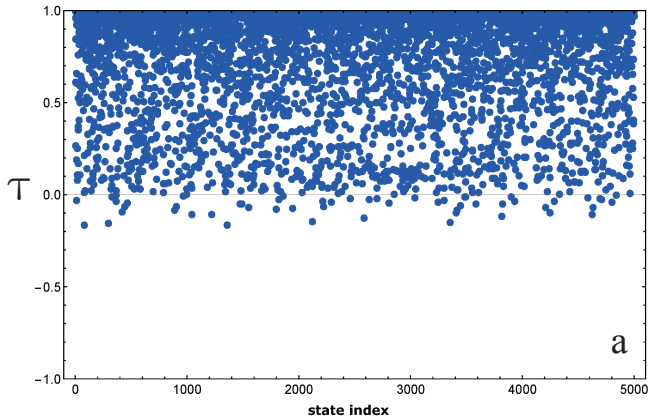}
\end{center}
\caption{(Color online) Kendall's $\tau$ coefficient between $F_{\text{c}}$ and $F_{\text{n}}$ against state index in randomly selected sample of $\tilde{M}=5\times 10^{3}$ initial states. Each initial state is sent through the amplitude damping channel for $M=2\times 10^{2}$ different values of $p$ uniformly distributed in $[0, 1]$ and a $M$-set of paired observations of quantities $F_{\text{c}}$ and $F_{\text{n}}$ is produced. The $\tau$ coefficient is plotted for the produced $M$-set of paired observations.}
\label{fig:FcFn}
\end{figure}

Note that unlike the special case discuss with a local bit-phase flip channel at the beginning of this section, the final states after the channels in Eqs. (\ref{Amplitude damping}-\ref{Phase flip}) turn to be two-qubit mixed states in general. Thus, 
as pointed out in sec. \ref{Negativity} based on Ref. \cite{miranowicz2004}, concurrence and negativity do not necessarily coincide for the final states and, in fact, they independently measure the entanglement of final states in latter cases. 
This independency becomes more clear in the Kendall's tau coefficient analysis between $F_{\text{c}}$ and $F_{\text{n}}$ through, for example, amplitude damping channel in Fig. \ref{fig:FcFn}.  The figure shows that the Kendall's tau coefficient between $F_{\text{c}}$ and $F_{\text{n}}$ vanishes for a number of paired observations as well as that the absolute value of Kendall's tau coefficient is not one for the most of paired observations.
This approves that $F_{\text{c}}$ and $F_{\text{n}}$ and consequently concurrence and negativity are independent quantities with different entangling aspects. 
Therefore, an ordinal correlation between $F_{\text{e}}$ and $F_{\text{c}}$ would be independent of an ordinal correlation between $F_{\text{e}}$ and $F_{\text{n}}$.
Below, we continue our analyses for $F_{\text{c}}$ and $F_{\text{n}}$ separately. 

\begin{figure}[H]
\begin{center}
\includegraphics[width=90mm]{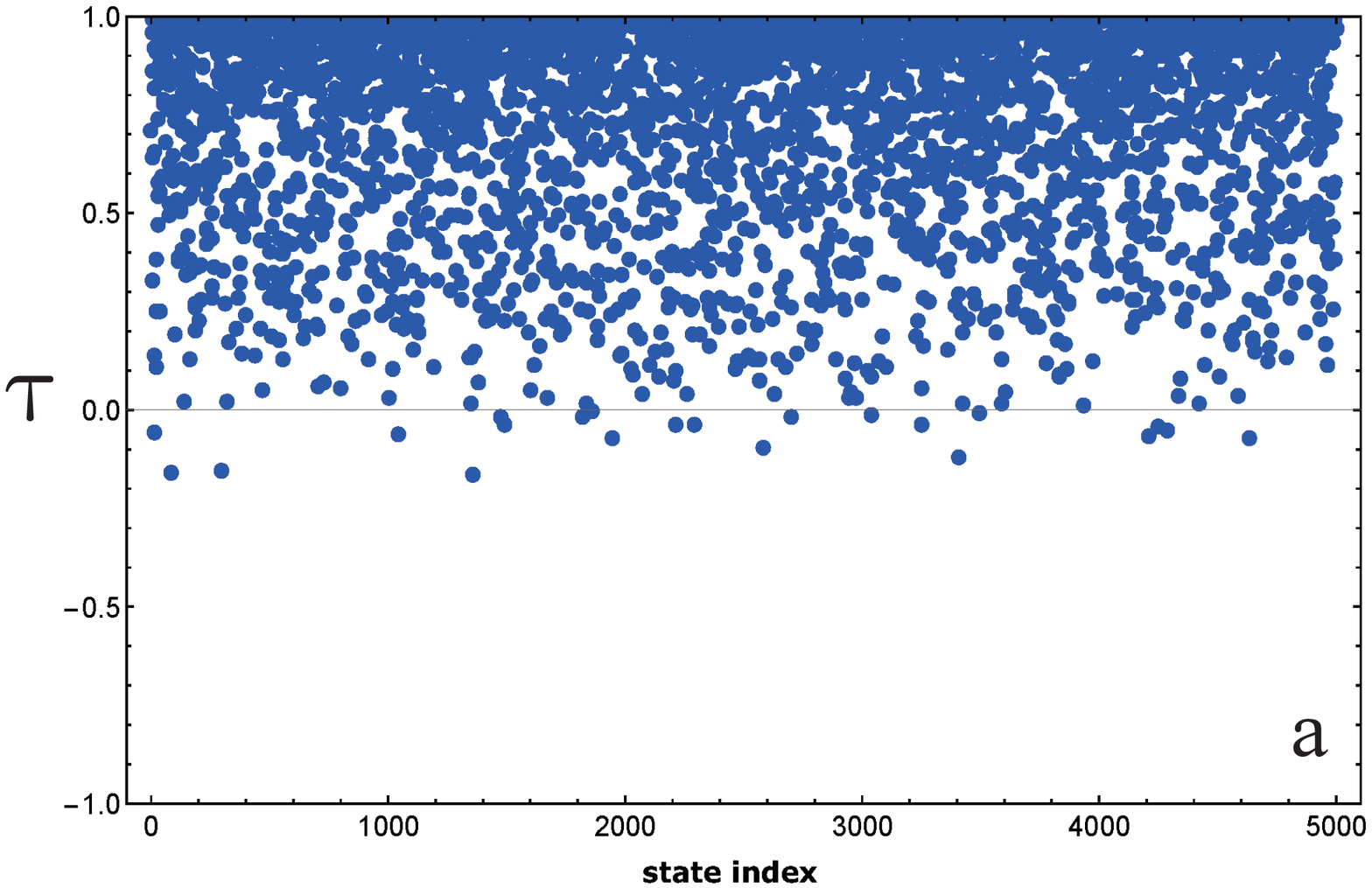}
\includegraphics[width=90mm]{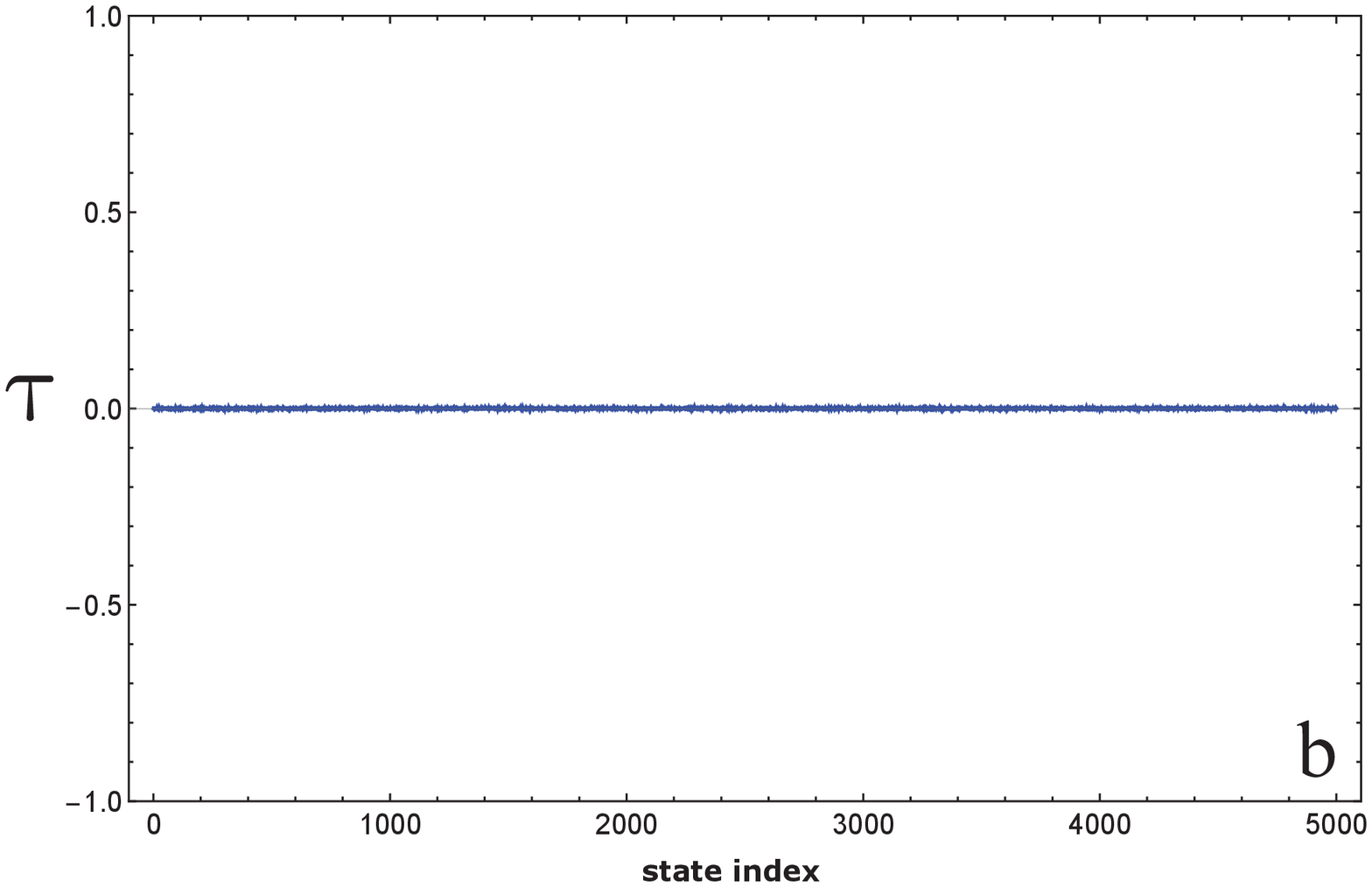}
\includegraphics[width=90mm]{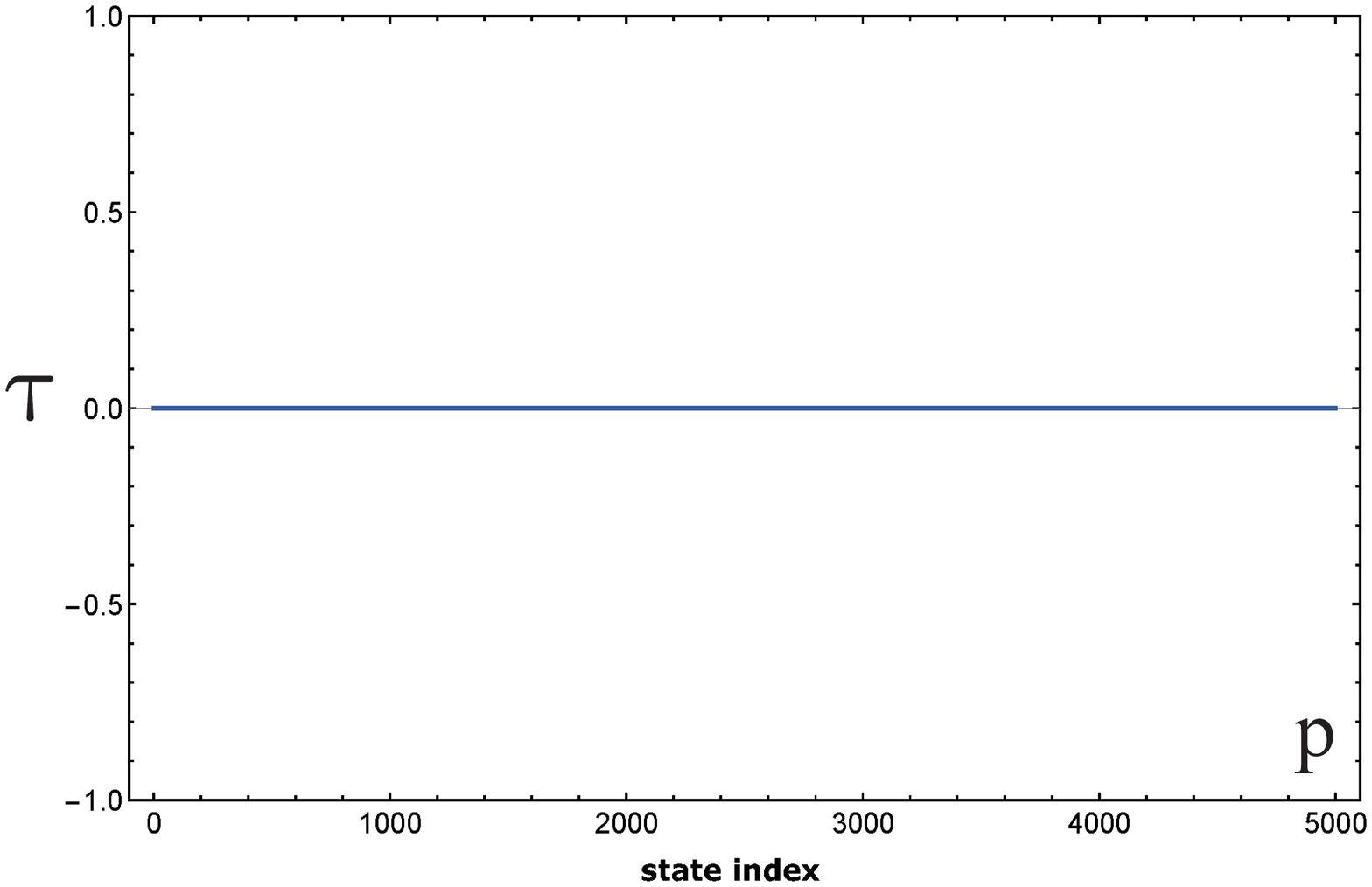}
\end{center}
\caption{(Color online). Kendall's $\tau$ coefficient between $F_{\text{e}}$ and $F_{\text{c}}$ against state index in randomly selected sample of $\tilde{M}=5\times 10^{3}$ initial states. The ''a'', ''b'' and ''p'' panels, respectively, correspond to ''amplitude damping'', ''bit flip'' and ''phase flip'' channels. Each initial state is sent through the given quantum channel for $M=2\times 10^{2}$ different values of $p$ uniformly distributed in $[0, 1]$ and a $M$-set of paired observations of quantities $F_{\text{e}}$ and $F_{\text{c}}$ is produced. The $\tau$ coefficient is plotted for the produced $M$-set of paired observations.}
\label{fig:FeFc}
\end{figure}

Fig. \ref{fig:FeFc} illustrates our analysis of the Kendall's tau coefficient for the pair $F_{\text{e}}$ and $F_{\text{c}}$ in the three quantum channels. As seen from the figure, not only the absolute value of Kendall's tau coefficient is not one in the most of the cases but it is zero in many cases particularly in the bit flip and phase flip channels, where we just get zero Kendall's tau coefficient for each initial state. This provide a solid evidence for the entanglement fidelity $F_{\text{e}}$ to be independent of the fidelity $F_{\text{c}}$. Therefore, the entangling aspect of the entanglement fidelity $F_{\text{e}}$ is not of type concurrence. Consequently, the $F_{\text{e}}$ is independent of $F_{\text{ef}}$ and the entangling aspect of $F_{\text{e}}$ is not of type entanglement of formation either. 

\begin{figure}[H]
\begin{center}
\includegraphics[width=90mm]{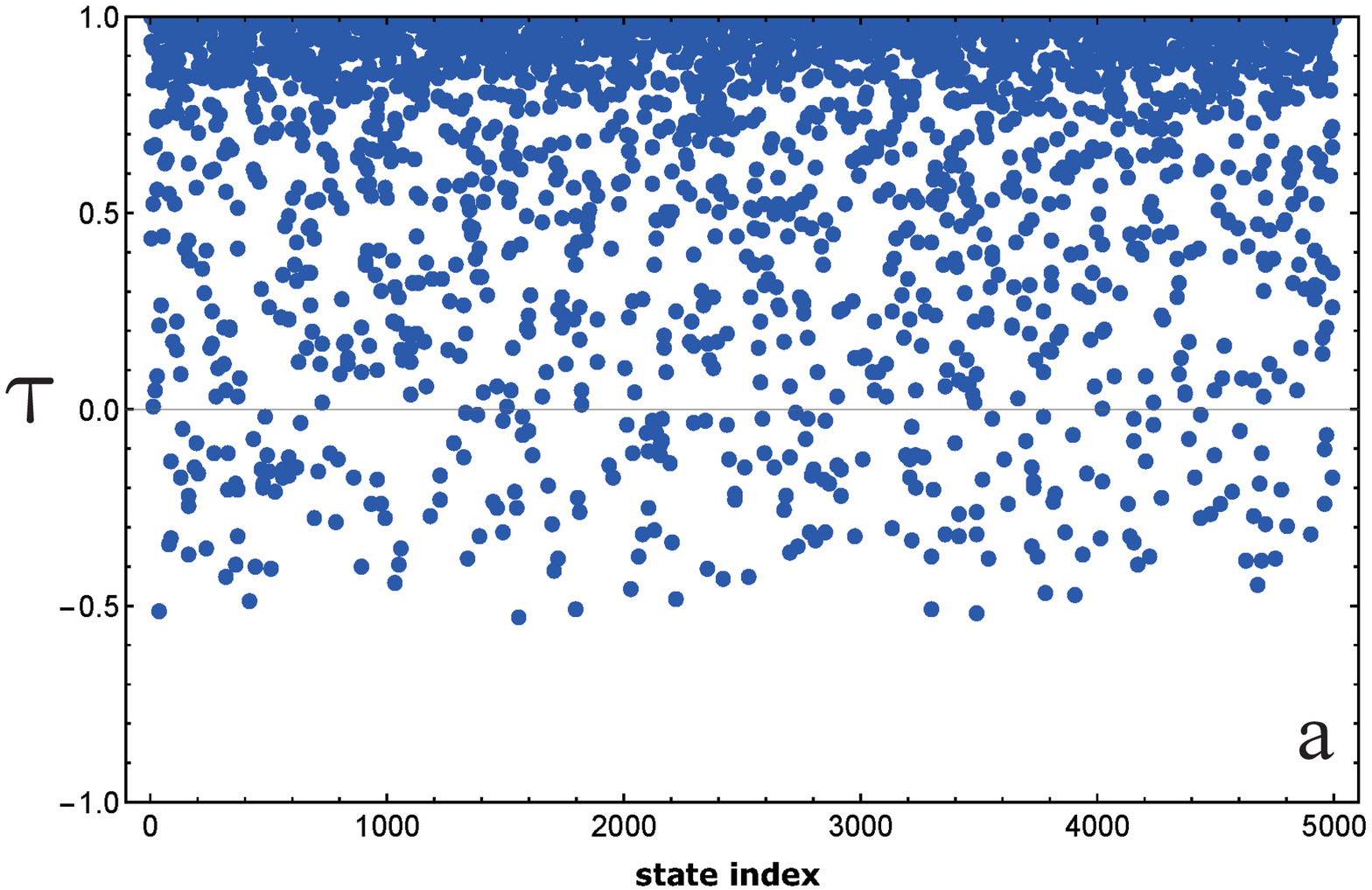}
\includegraphics[width=90mm]{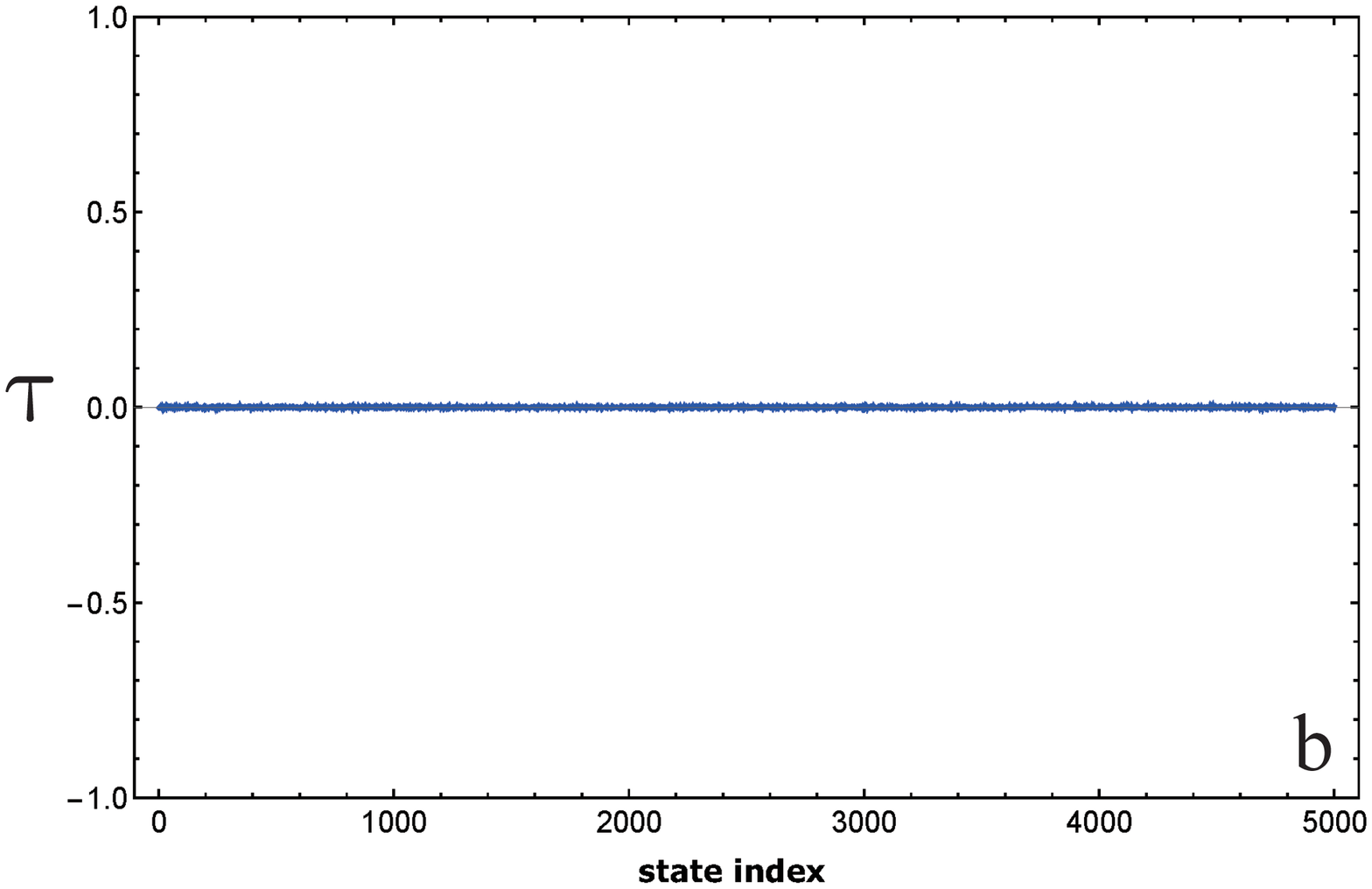}
\includegraphics[width=90mm]{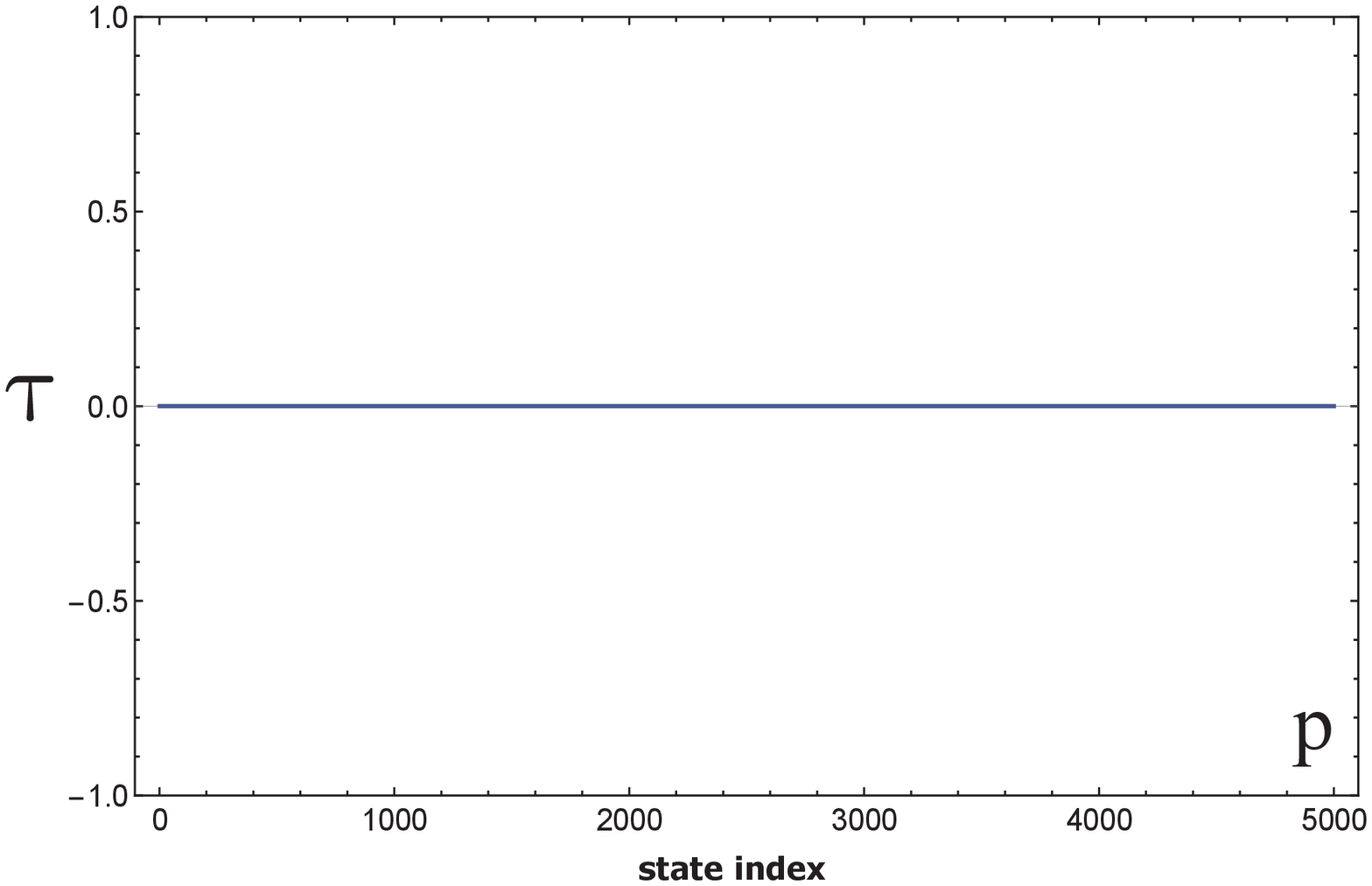}
\end{center}
\caption{(Color online). Kendalls $\tau$ coefficient between $F_{\text{e}}$ and $F_{\text{n}}$ against state index in randomly selected sample of $\tilde{M}=5\times 10^{3}$ initial states. The ''a'', ''b'' and ''p'' panels, respectively, correspond to ''amplitude damping'', ''bit flip'' and ''phase flip'' channels. Each initial state is sent through the given quantum channel for $M=2\times 10^{2}$ different values of $p$ uniformly distributed in $[0, 1]$ and a $M$-set of paired observations of quantities $F_{\text{e}}$ and $F_{\text{n}}$ is produced. The $\tau$ coefficient is plotted for the produced $M$-set of paired observations.}
\label{fig:FeFn}
\end{figure}

In Fig. \ref{fig:FeFn} we evaluate the Kendall's tau coefficient for the pair $F_{\text{e}}$ and $F_{\text{n}}$ in the three quantum channels. Similarly, we see that in the most of the cases the absolute value of Kendall's tau coefficient between $F_{\text{e}}$ and $F_{\text{n}}$ is not one and even in many cases is zero.  In the bit flip and phase flip channels, we notice that the Kendall's tau coefficient totally vanishes for the whole sample of initial states. Therefore, the entanglement fidelities $F_{\text{e}}$ and $F_{\text{n}}$ behave differently and indeed are independent from each other in principle. In a word, the entangling aspect of the entanglement fidelity $F_{\text{e}}$ can not be of type negativity.

We conclude this section with some remarks. First, the entanglement fidelities $F_{\text{e}}$, $F_{\text{c}}$ and $F_{\text{n}}$ are mutually independent quantities and therefore they cannot refer to the same entangling aspect of a quantum system. 
Second, there are cases where the fidelity $F_{\text{e}}$ fails to detect entanglement preservation or changes through quantum channels.
Third, each measure of entanglement defines individually its own entanglement fidelity and thus we expect the fidelity $F_{\text{e}}$ to associate with a measure of entanglement if it possess any entangling aspect. 
If it is not entanglement of formation, concurrence or negativity then what is the entangling aspect of the entanglement fidelity $F_{\text{e}}$? All these express that the question 'what is the most relevant way to quantify the variation of entanglement in a quantum process?' need a proper reconsideration.

\section{Summary}
\label{Summary}

In summary, we investigated entangling aspect of the entanglement fidelity introduced previously in Refs. \cite{schumacher1996, nielsen1996, barnum1998}. 
We introduced fidelity type quantities associated with some measures of entanglement, namely, entanglement of formation, concurrence and negativity. We used the statistical tool known as Kendall rank correlation
coefficient to study mutual correlations among the newly defined entanglement fidelities based on measures of entanglement as well as the entanglement fidelity in Refs. \cite{schumacher1996, nielsen1996, barnum1998}. 
With the analysis of Kendall rank correlation coefficient in two-qubit quantum systems subjected to three different quantum processes, we showed that there are no ordinal correlations between the entanglement fidelity in Refs. \cite{schumacher1996, nielsen1996, barnum1998} and the three measures of entanglement and indeed they are independent. This confirms that entangling aspect of the entanglement fidelity introduced in Refs. \cite{schumacher1996, nielsen1996, barnum1998} is neither of type entanglement of formation and concurrence nor of type negativity. Moreover, we showed that, in some cases, the entanglement fidelity in Refs. \cite{schumacher1996, nielsen1996, barnum1998} fails to detect entanglement preservation or changes through quantum channels.
Our analyses demonstrate that each measure of entanglement defines merely its own entanglement fidelity and thus if the fidelity in Refs. \cite{schumacher1996, nielsen1996, barnum1998} has any entangling aspect, it should associated with a measure of entanglement.

The results for Kendall rank correlation coefficient suggest it would be important to clarify the relation between the concept of entanglement fidelity and measures of entanglement. This would further improve our understanding of the quantum entanglement and its important role in quantum information processing. It would also help shed light on which of different types of entanglement is the most accessible experimentally and, at the same time, the most amenable to external manipulation and in general the most relevant to quantum information processing.

\section{Acknowledgment }
This research was supported by Department of Applied Mathematics and Computer Science at University of Isfahan (Iran) and in part by a grant from IPM
through grant No. 98810042.


%
%

\bibliographystyle{spmpsci}      
\bibliography{EFs}   

\end{document}